\documentclass{IEEEcsmag}

\usepackage[colorlinks,urlcolor=blue,linkcolor=blue,citecolor=blue]{hyperref}

\PassOptionsToPackage{hyphens}{url}
\usepackage{hyperref}

\usepackage[none]{hyphenat}

\PassOptionsToPackage{hyphens}{url}\usepackage{hyperref}

\usepackage{upmath}

\usepackage[normalem]{ulem}

\usepackage{color,soul}

\usepackage{hyperref}
\usepackage[none]{hyphenat}

\jvol{XX}
\jnum{XX}
\paper{8}
\jmonth{September}
\jname{}
\pubyear{2023}

\usepackage[hyphens]{url}
\usepackage{hyperref}
\hypersetup{colorlinks=true,breaklinks=true}

\usepackage{xurl} 
\hypersetup{breaklinks=true} 

\usepackage{comment}

\newcommand{\chop}[1]{}

\begin{document}

\sloppy 


\title{Game Theory in  Distributed Systems Security: Foundations, Challenges, and Future Directions}

\author{Mustafa Abdallah$^1$, Saurabh Bagchi$^2$, Shaunak D. Bopardikar$^3$, Kevin Chan$^4$, Xing Gao$^5$, Murat Kantarcioglu$^6$, Congmiao Li$^7$, Peng Liu$^8$, Quanyan Zhu$^9$ \\}

\affil{Apart from the first two authors, all other authors are listed alphabetically. Corresponding author: Saurabh Bagchi (sbagchi$@$purdue.edu)}

\author{}
\affil{(1) Indiana University-Purdue University Indianapolis; (2) Purdue University; (3) Michigan State University; (4) Army Research Lab; (5) University of Delaware; (6) University of Texas at Dallas; (7) University of California at Irvine; (8) Pennsylvania State University; (9) New York University}



\markboth{}{Paper title}

\begin{abstract}
Many of our critical infrastructure systems and personal computing systems, that have a distributed structure, face increasing levels of attacks.  
There has been a vast research on both game theory and distributed system security to face these increasing attacks.
Therefore, we feel it is time to bring in  the rigorous reasoning from game theory advanced models to better secure such distributed systems. The distributed system security and the game theory technical communities can come together to effectively address this challenge of securing distributed systems. In this article, we lay out the foundations from each domain that we can build upon to achieve  a successful integration of game theory and distributed system security for better securing large-scale distributed systems. 
We then describe a set of research challenges for the community, organized into three categories --- analytical, systems, and integration challenges, each with ``short term" time horizon (2-3 years) and ``long term" (5-10 years) items. 
\end{abstract}

\maketitle

\chapterinitial{Introduction}

\begin{figure*}[t]
\centerline{\includegraphics[width=0.94\linewidth]{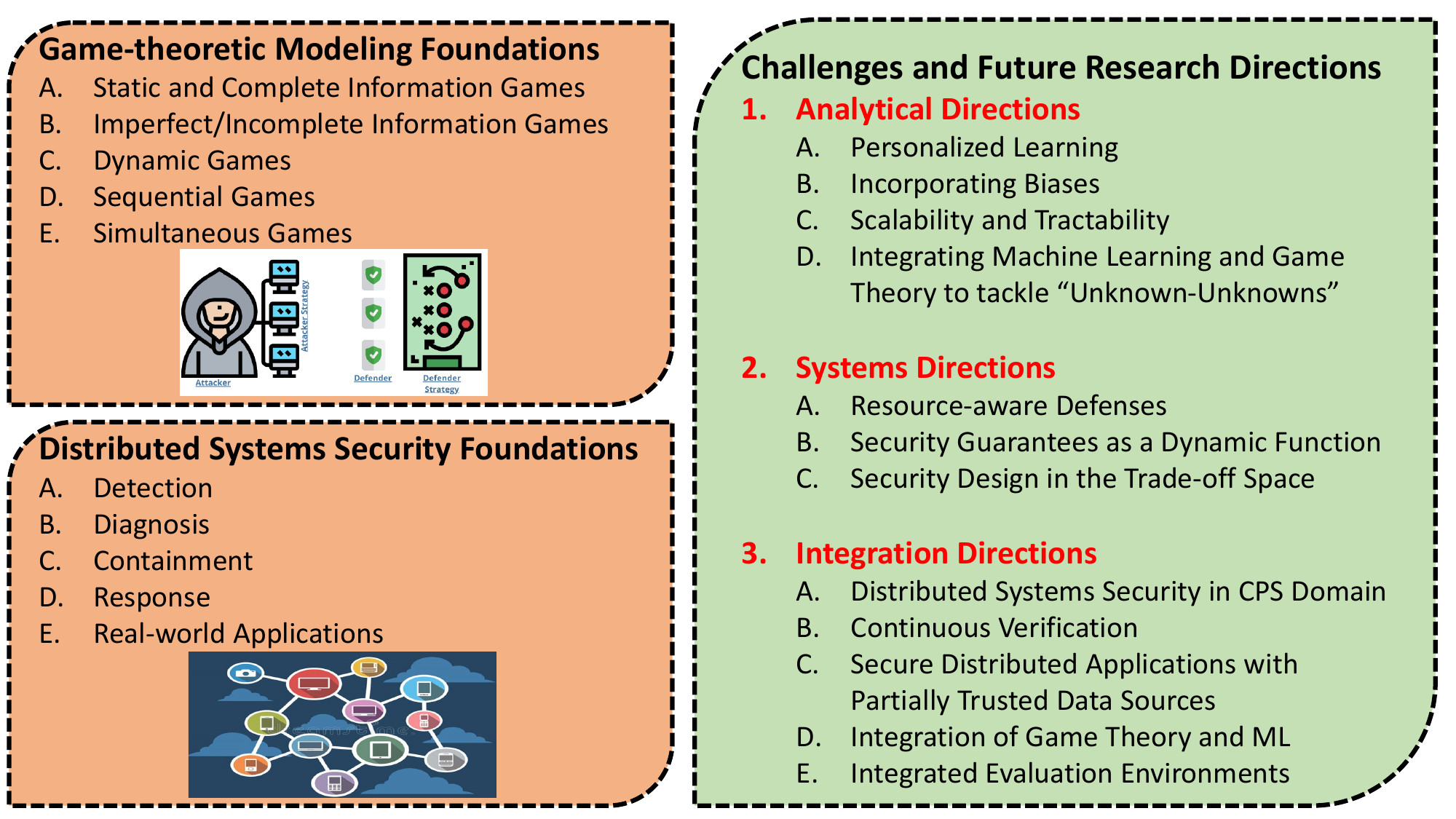}}
\vspace{-3mm}
\caption{An overview of the flow of this paper. We first show the main foundations for game-theory modeling, and distributed systems security. We then outline the research challenges and future directions that will need the integration of the advancements of the analytical side  and systems side for securing distributed systems.}
\label{fig:paper_flow}
\vspace{-2mm}
\end{figure*}

Today's distributed systems face sophisticated attacks from external adversaries where the attacker aims to breach specific critical assets within these systems. Such attacks pose a serious danger to large-scale critical infrastructure, such as, the massive supply chain attack on SolarWinds in 2020 and the Colonial Pipeline ransomware attack in  2021. Such attacks have motivated several attempts to improve the cyber security of these systems~\cite{humayed2017cyber}. 

In response to such attacks, there has been significant work in understanding vulnerabilities in large-scale distributed systems and putting technological patches to address specific classes of vulnerabilities. 
However, the works often lack understanding of the impact of cascading attacks or of mitigation techniques, on the security of the overall system. 
Due to the large legacy nature of many distributed infrastructures and budgetary constraints, a complete re-architecting and strengthening of the system is often impossible. 
Rather, rational decisions must be made to strengthen parts of the system, taking into account the risks and the interdependencies among the assets. In this context, significant research has investigated how to better secure these systems, with game-theoretical models receiving increasing attention. Such models have shown the power to capture the interactions of different players (strategic attackers and defenders) in different settings (see the survey~\cite{laszka2015survey}).

While researchers have studied static game theory extensively for several decades,  large-scale distributed systems present critical challenges that preclude the direct application of existing theory.
Specifically, there is a need for new techniques to account for both the interdependencies and the dynamical nature of the subsystems.
Furthermore, some of these dynamical subsystems may be complex in their own right (e.g., a perception system that employs multi-modal sensors) and may have the limitation of only being represented by simulation models. Thus, advanced game theory models can be proposed to better model attacker/defender realistic scenarios, where such modeling should be connected more to distributed system security to find new insights in securing distributed systems. 

This problem context leads to four overarching questions that form a starting point for enhancing usage of game theory for distributed system security.\vspace{-2mm}
\begin{enumerate}
    \item Can the security community extend traditional game theory to develop tractable analysis and design techniques that can be applied for securing {\em large-scale} and {\em interdependent} distributed systems?

    \item What are the main foundations in game theory and distributed system security literature that can help us to achieve such a goal of securing distributed systems?

    \item What are  advantages and disadvantages of different game-theoretic models when applied to distributed systems security?

    \item What are the main challenges and related research directions for integrating game theory  for securing distributed systems?\vspace{-1mm} 
\end{enumerate}

In this article, we present a proposed vision about answering these questions. In particular, we organize our paper as follows. We first lay out the foundations that the research community can build on when applying game-theory concepts in order to enhance distributed system security. We then present the main challenges for such a synthesis, which we categorize into (1) Analytical directions, (2) Systems directions, and (3) Integration directions. Figure~{\ref{fig:paper_flow}} provides the main flow of this paper.

\section{Foundations: Build on Them}

We have significant foundations on the topics of distributed systems security and game-theoretic security that we should build upon. Here we survey the notable foundations categorizing them into two --- game-theoretic security and distributed systems security.


\subsection{Game-theoretic Security}



There have been notable successes in developing and applying game theory for security of distributed systems~\cite{abdallah2020behavioral}. This has been used in the context of proactive or reactive and fixed or adaptive schemes. The most commonplace game-theoretic model for security is that of two-player games, where a single attacker attempts to compromise a system controlled by a single defender. The game-theoretic models have been further used to study the interaction between (multiple) defenders and (multiple) attackers (e.g., analyzing distributed denial of service attacks (DDoS), and security of cyber-physical systems).  The literature on game-theoretic models (and their unique differences) for different security scenarios can be categorized as follows.

\noindent \textbf{Static and Complete Information Games:}
The static and complete information two-player games are benchmark security models that capture the incentives or objectives of the players as well as their constraints. The game assumes that the players have a common knowledge of the game, and that it does not change over time.  The Nash equilibrium of the game can be interpreted as the outcome of repeated plays between the two players or the consequence of homogeneous pairwise interactions of a large population. 
The analysis this class of games provides a quantitative approach to assess risks and to design mitigating mechanisms. FlipIt games and Blotto games  \cite{roberson2006colonel} 
are two notable games that have been widely used in understanding the competitive scenarios of resource takeover and subjugation in cyber-physical systems and military applications. The Nash equilibrium is the traditional concept capturing efficient solution(s) of complete information games. 
For instance, Nash equilibrium has helped understand botnet defenses~\cite{botnet-defense}. In particular, this line of work has provided a comprehensive game theoretical framework that models the interaction between
the botnet herder and the defender group (network/computer users). The Nash equilibrium showed the effectiveness of available defense strategies and control/strategy switching thresholds, specified as rates of infection.
The two Nash equilibria obtained are either (1) the defender group defends at maximum level while the botnet herder exerts an intermediate constant intensity attack effort or (2) the defender group applies an intermediate constant intensity defense effort while the botnet herder attacks at full power. 
This model also showed that integrating game-theoretical analysis with SIS epidemic models could be useful in understanding system behavior during botnet attacks. 

Overall, although complete information game-theoretic models for security games enable proactive security planning and predicting worst-case attack scenarios on these distributed systems, actual conflicts are dynamic and involve incomplete information for one or both players, that are discussed next.

\noindent \textbf{Imperfect/Incomplete Information Games:}
%
These are games in which at least one player (defender or attacker) does not have complete information. This may be due to lacking complete knowledge of the system or to imprecise sensing. To analyze multi-stage, multi-host attacks that may be launched on networks, a defender needs to model long sequences of actions that can circumvent the system defenses \cite{alpcan2010}. These actions lead to policy spaces that grow exponentially with the number of attack stages, especially under partial/imperfect information. Monte Carlo sampling can confine the search to a decision tree of reduced size by guessing the other
  player's moves, and then using a conventional minmax search to
  determine the best strategy~\cite{sandholm2015abstraction}. 
  
  One promising line of work for games with imperfect/partial information is the use of deception (see a recent survey~\cite{zhu2021survey}). The key idea is for one or both players to synthesize new actions/policies that leverage limitations induced by the belief of the opponent. Notable classes of problems that fall within this class are \emph{signaling games} that model information corruption, \emph{Bayesian games} that model uncertainty in opponent's type/cost, \emph{asymmetric constraints} that enforce stealth and partially observable stochastic games. 
  Akin to general imperfect information games, the complexity of solving deception problems grows exponentially with the number of stages, beliefs, and actions. 
  
  Drawing inspiration from robust optimization, the application of randomized sampling methods has proven effective in computing policies that are robust security measures against adversaries employing randomized strategies~\cite{bopardikar2013randomized}. 
  These methods utilize randomized sampling techniques to explore the strategy space to choose effective strategies with high confidence. 


Overall, leveraging incomplete information game-theoretic
models for distributed systems security captures the uncertainty about the adversary's actions and payoffs, along with the actions of other stakeholders which can give more accurate quantitative estimation of the security level of the distributed system. However, if the players information evolves over time, then they are more effectively modeled as dynamic games.
  

\noindent \textbf{Dynamic Games:}
These are games in which the information, player actions or the payoffs vary over time. One promising line of work has been in leveraging \emph{Reinforcement Learning (RL)}. Examples of such usage of RL are malicious falsification of cost signals that is used to mislead agent policy~\cite{huang2019deceptive}. Another example is applying RL and infinite-horizon Semi-Markov Decision Process (SMDP) to characterize a stochastic transition and sojourn time of attackers in a honeynet.  Another line of work is to model distributed systems using Hybrid Input-Output Automaton (HIOA). This can help in characterizing the continuous time evolution of the security game.

In contrast to static game-theoretic models, these dynamic games capture the realistic evolution of vulnerabilities, and adversary actions which can lead to effective usage of learning-based techniques for guiding human (or automated) decision-making towards better security policies for securing current distributed systems that have such dynamic nature. However, if there is a natural order in the conflicts that requires one player to play first or if the actions of both players are not visible to each other until a specified time, then such situations are more effective modeled as sequential or simultaneous games, respectively, as discussed next.

\noindent \textbf{Sequential Games:} Game theory for security has been found to be tractable when considering sequential attacks, through {\it Stackelberg security games}. In these games, the defender moves first and allocates her resources to the assets under her ownership. Then, the attacker can observe the allocations made by the defender to each asset, after which he targets part (or all) of the assets. Such games may incorporate real-time observations and consideration of non-myopic players. In reality, many such games may be partially observable as the actions of a player may not be visible to other players (e.g., an attacker may conceal her steps). 

There have been several applications that have benefited from Stackelberg security games for distributed systems, as diverse as countering Man-In-The-Middle (MITM) attacks and 
screening airport passengers throughout the USA~\cite{sinha2018stackelberg}. The sequential order in these games also identify realistic cases where adversary attacks distributed systems (in firms or government infrastructure) after the security decision-makers invest in securing these systems. 

\noindent \textbf{Simultaneous Games:} 
A particular class of simultaneous move games involving attackers and defenders (where the players have to choose their strategies simultaneously, without first observing what the other player has done) has been studied in various contexts.  For example, the Colonel Blotto game 
is a useful framework to model the allocation of a given amount of resources on different potential targets (e.g., battlefields) between the attacker and the defender. Specifically, \cite{7943422} proposed a solution for the heterogeneous Colonel Blotto game with asymmetric players (i.e., with different resources) and with many battlefields that can have different values. While Colonel Blotto games typically involve deterministic success functions (where the player with the higher investment on a node wins that node), other work has studied cases where the win probability for each player is a probabilistic (and continuous) function of the investments by each player. 
Overall, simultaneous-move games arise in military-based distributed system security applications. Furthermore, simultaneous-move games may be a better way to model real-world situations in which attackers may choose to act without acquiring costly information about
the defense security strategy, particularly if security measures are difficult to
observe (e.g., undercover officers, strong privacy measures, and non-available insiders).

\noindent {\bf Advanced Games Examples for Security of Distributed Systems:}  Game theoretic models have also been used to study DDoS attacks, critical infrastructure security, censorship-resilient proxy distribution, wireless network security, and protecting computer networks from cascade attacks (see the survey~\cite{laszka2015survey}). Further,~\cite{abdallah2022tasharok} studied mechanism design to incentivize defenders towards beneficial security investments in distributed systems.

\begin{figure*}[t]
\centerline
{\includegraphics[width=\linewidth]{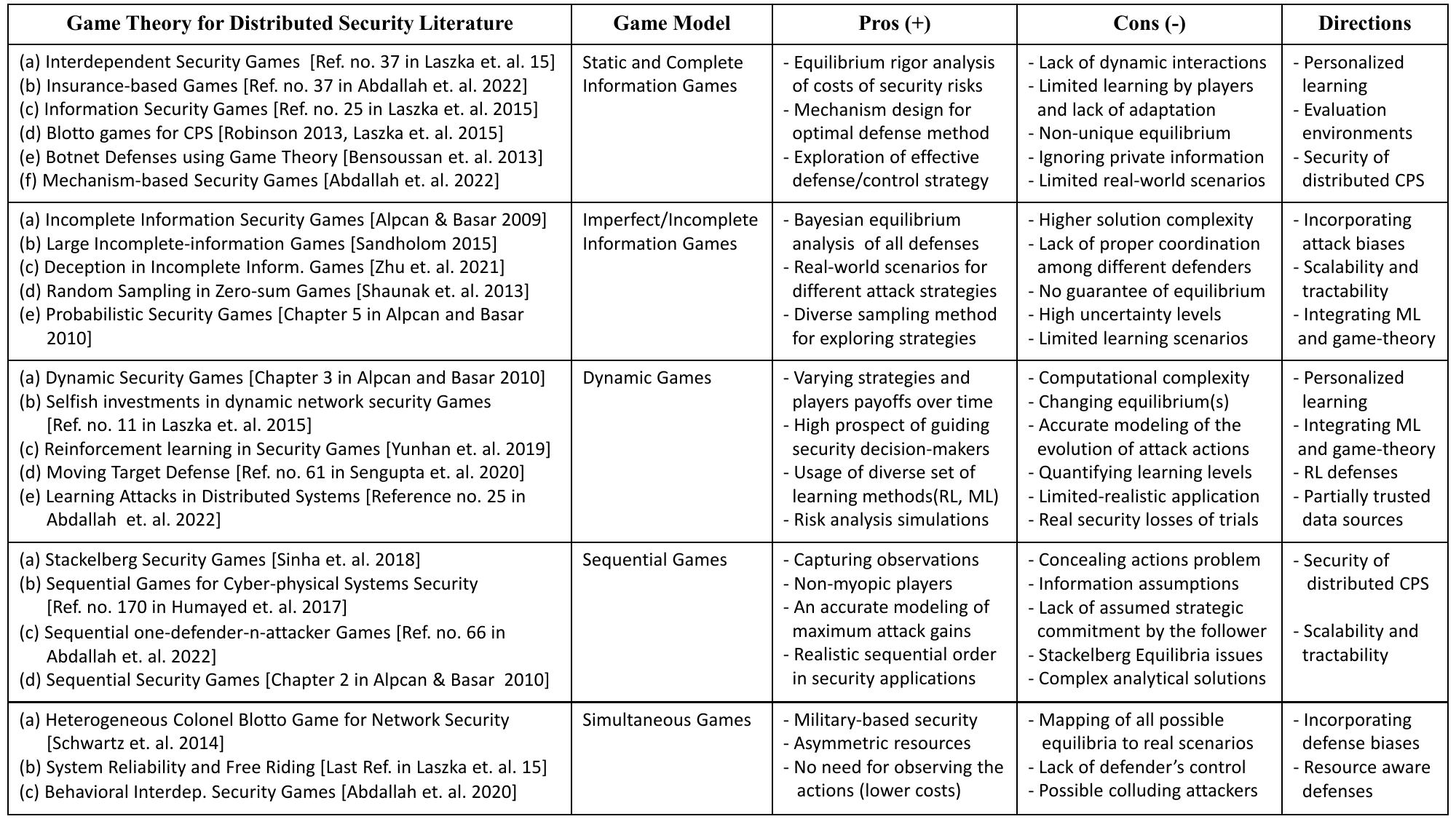}}
\vspace{-2mm}
\caption{A summary of the relevant literature of game-theoretic models for distributed systems security. We show the pros and cons of various game-theoretic models as applied to distributed systems security and prospective usage of each model in the different research directions outlined in our vision.}
\label{fig:review_pros_cons_direcs_game}
\vspace{-3mm}
\end{figure*}

\textbf{Summary of Game-theory Literature on Distributed Security:} Figure~\ref{fig:review_pros_cons_direcs_game} provides an overview of the literature on game-theoretic models for distributed systems security. We highlight the advantages and disadvantages of different game-theoretic models when applied to distributed systems security and discuss the potential applications of each model in the various research directions outlined in our vision.


\subsection{Distributed Systems Security}
One way to organize the foundations that have been developed here is through each step of the workflow for distributed systems security, namely, detection, diagnosis, and containment and response. 

\noindent {\bf Detection:} This is a mature area of work in which there is influential work on collaborative intrusion detection using multiple sensors placed in a distributed system. This line of work has contributed algorithms to determine where to place the sensors and how to integrate outputs from multiple sensors to devise an integrated decision on the detection of an attack. A survey work on this topic is~\cite{vasilomanolakis2015taxonomy}. This area saw some of the early applications of ML to security.
In the context of game theory, game-theoretic analysis has helped develop various intrusion detection systems for distributed systems to increase detection accuracy with reduced cost.
Game-theoretical approach has also been used for mitigating edge DDoS attacks~\cite{laszka2015survey}.


\noindent {\bf Diagnosis:} It has contributed algorithms to identify the root cause of the attack. This was initially substantially rule-based, of the form, if metric A $>$ threshold $\tau_1$ and B $<$ threshold $\tau_2$, then A is the root cause~\cite{khanna2007automated}. Later, foundational work was done on this topic on using ML, such as causal theory~\cite{wang2018towards}. 
One significant challenge that has been successfully addressed is how to maintain effective diagnosis capabilities in security algorithms when the interactions and connections between elements within a distributed system are constantly changing.

\noindent {\bf Containment and Response:} This concept has had notable success in the topic of Moving Target Defense, which seeks to change some parameters of the defended system such as IP address to thwart an adversary. 
This can be done proactively as a preventive measure in response to a detected threat~\cite{sengupta2020survey}. 

The integration of game analysis for critical infrastructure protection has proven highly successful, effectively encompassing containment, response, and recovery measures. 

\noindent {\bf Applications: CPS and Critical Infrastructure}

Security games have played a crucial role in addressing the resilience and interdependence of critical infrastructures, including our nation's legacy cyber-physical systems such as power grids~\cite{abdallah2022tasharok}, transportation~\cite{alpcan2010}, and manufacturing systems. With the increasing connectivity of these systems, they face a larger attack surface. The application of game-theoretic methods is vital in developing strategic mechanisms for detection, diagnosis, containment, and response, ensuring their resilience. 

Having outlined the foundational aspects of game-theoretic models and distributed systems security, we now turn our attention to the key challenges faced by the research community and provide prospective research directions.



\section{Challenges Ahead}

\begin{figure*}[t]
\centerline
{\includegraphics[width=0.99\linewidth]{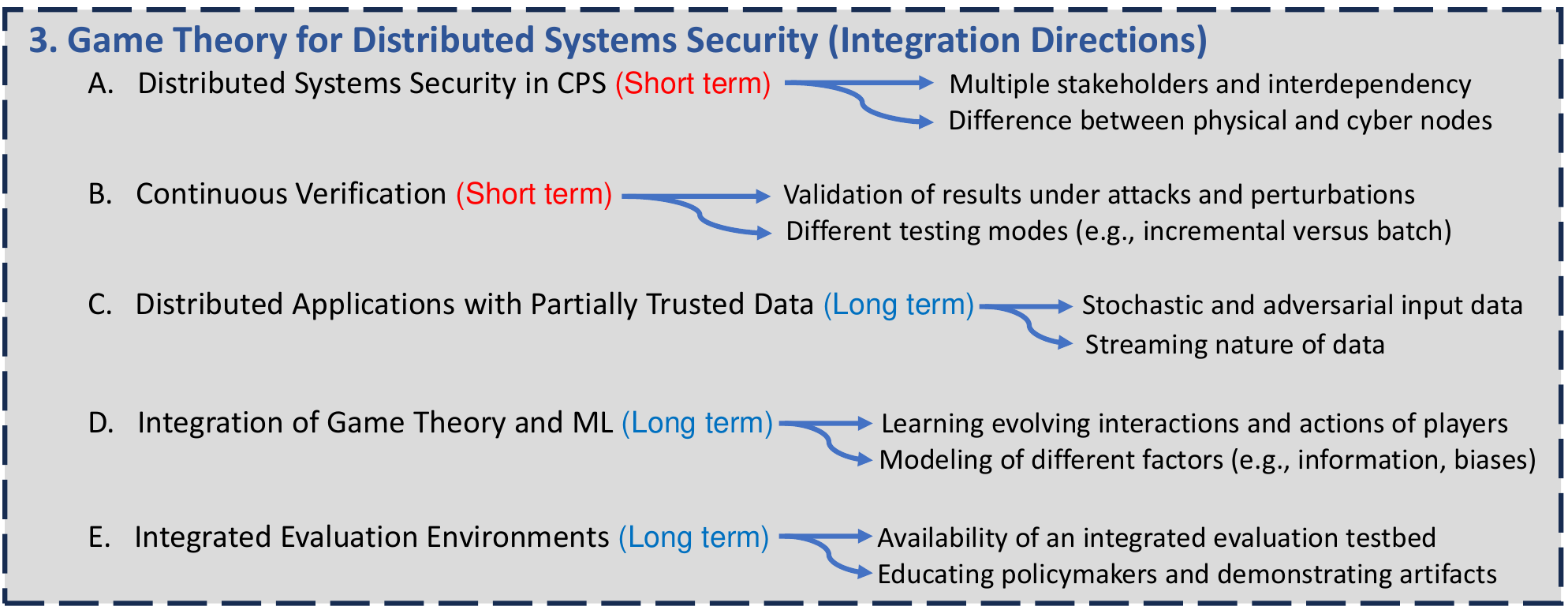}}
\vspace{-2mm}
\caption{A timeline overview of research challenges and future research directions for both of analytical side and systems side (upper part), along with possible research directions of integrating both sides (lower part).}
\label{fig:timeline}
\vspace{-1mm}
\end{figure*}

Here we summarize the technical challenges in order to improve the security landscape of distributed systems. We structure our discussion into challenges that are on the {\bf analytical directions}, {\bf systems directions}, and those that involve combination of the two called, {\bf integration directions}. The orthogonal dimension on which we structure these challenges is the time horizon to solve them, with short term indicating 2-3 years and long term indicating 5-10 years. Figure~{\ref{fig:timeline}} summarizes main challenges and future research directions for such integration.\vspace{-2mm} 

\subsection{Analytical Directions}
\label{sec:analytical-directions}


\noindent {\bf Personalized learning:} \textcolor{red}{(Short time horizon)} Different actors (say different defenders and adversaries) learn differently (as in stochastic learning) and at different rates. The learning happens for human actors as well as for machines (in a ML context). This learning can build on literature on Dynamic Games, discussed in Foundations. Another form of heterogeneity that comes in is asymmetric capabilities among the various players, e.g., some defenders have access to assets with a trusted hardware environment, like ARM's TrustZone or Intel's SGX.     The personalization of the learning must also be able to accommodate partial cooperation (among defenders) or partial collusion (among adversaries) (building on literature on Incomplete Information Games, discussed in Foundations), in addition to the formulation of complete cooperation or collusion (building on  Complete Information Games, from Foundations).  
    
\noindent {\bf Incorporating biases and incomplete information:} \textcolor{red}{(Short time horizon)} 
    For the learning, one has to incorporate incomplete information sharing among the actors (building on prior literature on Incomplete Information Games, discussed in Foundations). For the human learning, one must also incorporate cognitive biases among the human players. Behavioral economics has shown that humans consistently deviate from these classical models of decision-making, e.g., humans perceive gains, losses and probabilities in a skewed, nonlinear manner~\cite{kahneman1979prospect}. We have done some nascent work applying behavioral game theory to security of distributed interdependent systems~\cite{abdallah2022tasharok}. The main point here is to explore the effects of such behavioral biases on security policies of human decision-makers, and their effect on securing distributed systems. One example in related literature is that behavioral human security decision-makers may allocate part of their limited security resources on non-critical parts of the distributed system. The key insight is that one can provide appropriate incentives to reduce the biases and encourage cooperation among even biased defenders. Such bias incorporation would build on prior Simultaneous Security Games for distributed systems security from Foundations. Such cooperation is in general a more secure strategy than independent decision making.  
    A related theme here is trust building among human agents in multi-agent learning. This is to counter the natural tendency for each player to explore and exploit their own strategy spaces (e.g., see the reference (a) in Dynamics Games in Figure~\ref{fig:review_pros_cons_direcs_game}).

    \noindent {\bf Scalability and tractability:} \textcolor{blue}{(Long time horizon)} A well-known challenge with applying game theoretic formulation to security of distributed systems is the scalability and the tractability of the solution (particularly for Imperfect Information Games and Sequential Games discussed in Foundations). Scalability implies scaling to the large numbers of actors or large amounts of data or large volumes of interaction among the actors. Tractability implies being able to handle realistic attack models or realistic workloads incident on the protected system. To ease this challenge, we must develop sound approximations of the game-theoretic formulation, e.g., leveraging sampling techniques discussed in the foundation part under ``Imperfect/Incomplete Information Games''. This should allow one to produce bounds for best-case/worst-case outcomes. As an example, one can use scalable techniques from epidemic theory to analyze the effect of cascading attacks while accommodating the case of large numbers of players.

    \noindent {\bf Integrating Machine Learning and Game theory to tackle ``Unknown-Unknowns":} 
    \textcolor{blue}{(Long time horizon)} The game theoretic formulations are often rigidly deterministic in nature, e.g., a specific deterministic action is coded in for a particular state. The open question is can machine learning be integrated with game theory and thus incorporate stochastic behavior. The best candidate game for that direction is Dynamic Games, discussed in Foundations. This is important as failures and attacks are inherently stochastic in nature. 
    The core challenge here is that machine learning methods can achieve accurate predictions only if they have been trained with the right set of examples. Security problems such as zero-day attacks remain extremely challenging because of the lack of appropriate types and numbers of examples until the current time instant. The key question to ask here is how can a defense scheme be resilient to unanticipated attacks (also known as ``black-swan events"). 
    

\subsection{Systems Directions}

\noindent {\bf Resource-aware defenses:} \textcolor{red}{(Short time horizon)} Different nodes have different capabilities and available resources and the system should be able to calibrate the defense mechanism using node-specific attributes. Some of these node-specific attributes will be static and unchanging, such as, the intrinsic hardware capability of the node (which can be captured efficiently using Static and Complete Information Games discussed in Foundations). In contrast, some attributes will be dynamic, such as, the current battery level on the node (which is better captured using Dynamic Games discussed in Foundations). The cost of an attack may also be varying, e.g., the cost to corrupt data may be higher if there is some security protection overlaid on the data. 

    Compared with traditional defense mechanisms which could be slow due to the lack of awareness of available resources and capabilities for different nodes, game-theoretic approaches can better allocate limited resources of each node to balance the defense tasks and take timely action. This is particularly important for critical infrastructure security. 

\noindent {\bf Security guarantees as a dynamic function:} \textcolor{red}{(Short time horizon)} The security guarantees may, under certain situations, be a function of the current system state. As an example, under this regime, the guarantees could be a function of the number and the capability of attackers and defenders rather than an absolute. Thus, the security guarantees that the system can provide are a dynamic property varying with the system state. For example, hardware degrades and the software ecosystem changes over time. The guarantees could also be a function of the level of collusion among attackers, e.g., non Byzantine or Byzantine attackers (which can be modeled by Imperfect Information Games from Foundations).

\noindent {\bf Designing for security in the tradeoff space:} \textcolor{blue}{(Long time horizon)} A radical design principle would be to design for security in the tradeoff space between security on the one hand and performance and resource usage on the other. For example, the security design may use hardware-level virtualization, when available, rather than (software) containers, the former providing stronger isolation and greater protection against side channel attacks. This direction can build on the literature on Containment and Response, discussed in Foundations. Suppose one can design specialized functions, say, specialized to the resource available at a node. In that case, this has the added benefit of reducing the attack surface, making debugging easier, and reducing consumed resources. 
    The security guarantees must be clearly delineated as a function of the performance and the resource usage, so that the end user can understand the guarantees she is getting. Alternately, in case of automatic composition with other software packages, it becomes clear what security guarantees are in effect in the composed system.  



\subsection{Integration Directions}

\noindent {\bf Security of distributed systems in CPS domain:} \textcolor{red}{(Short time horizon)} To secure Cyber-Physical Systems (CPS), there are several unique aspects that we need to consider. 
    These are prototypical interdependent distributed systems, often with multiple stakeholders as owners. The nodes are embedded in the physical environment and are subject to environmental effects, which contribute to the difficulty of securing them. For example, it is often difficult to tell apart a node malfunction due to environmental effects from a node compromise. Further, some parts of the system are opaque to defenders, as they are developed by an external party. Consequently, security mechanisms that rely on (fine-grained) observability of the events happening in the software stack are out of bounds. In this context, Sequential Games and Complete Information Games have a strong foundation to build on for these challenges (see related works in Figure~\ref{fig:review_pros_cons_direcs_game}). 

\noindent {\bf Continuous verification:} \textcolor{red}{(Short time horizon)}
    This topic needs to answer the question: are our models and practical software instantiations generating valid results even under attacks and perturbations? This should happen continually rather than in batch mode as is typically done today, when verification is used at all. The continuous verification should happen as the system processes inputs and generates outputs during its operation. This could use {\em sparse} human feedback online, i.e., without putting undue cognitive burden on the user. Existing methods for incremental verification/testing would be useful for this challenge~\cite{yao2021distai}, as would be recent progress on verification of highly non-linear ML models~\cite{yao2021distai}.
    This continuous verification would be based on Dynamic Games that can efficiently model this progressive verification setup.

    \noindent {\bf Secure distributed applications with partially trusted data sources:} \textcolor{blue}{(Long time horizon)}
    The overarching question that we need to answer is whether we can build secure distributed applications when the input data is only partially trusted. 
    This is particularly important for the significant class of systems that are stochastic and data dependent in nature. The data may be streaming, rather than at rest, adding to the challenge of verifying the data. The nodes comprising the distributed system are heterogeneous (as argued in an earlier item) in terms of resources, including secure hardware and access to trustworthy data sources. Finally, the trust in data is a dynamic property, increasing say when there has been successful validation of data and decreasing when there is a detected attack. This challenge can benefit from work on causal reasoning in dynamic systems as outlined under Foundations.

    \noindent {\bf Integration of game theory and machine learning:} \textcolor{blue}{(Long time horizon)}
    The grand open question here is can we, in a principled manner, integrate game theory and machine learning to secure distributed systems? In such integration, we must be 
    cognizant that there can be multiple players (tens to hundreds) in terms of attackers and defenders. The interactions and actions may evolve over time~\cite{zhu2021survey}, necessitating learning, rather than static spaces for actions and rewards, as is typical today. Dynamic Games is the best candidate to tackle such intergration challenges.

    There may also be partial information sharing among defenders, asymmetric information between attackers and defenders (captured by Imperfect Information Games from Foundations), and cognitive biases among players. A subset of these factors may be relevant in a specific application context, but the framework and the algorithms should be able to encompass them. 
    One direction could be to leverage ML models to learn the approximate utility functions of the participants by analyzing the past behavior data. ML models can also be used to learn approximate best response strategies when solving the game theoretical model is intractable (e.g.,~\cite{WuKKV21}).

\noindent {\bf Using game theory to gain better understanding of reinforcement learning defenses:} \textcolor{blue}{(Long time horizon)} 
Reinforcement learning (RL) techniques are being increasingly used to 
respond to continuous probes, e.g., heartbeat requests to exploit some vulnerability like the HeartBleed OpenSSL vulnerability and lateral movement attempts. 
Hence, the following research questions become important: 
Whether the ``game plays" between the attacker and the 
corresponding RL agent converge to a certain notion of equilibrium? 
And if so, how soon? 
A clear answer to these questions, that will build on Dynamic Games from game-theory foundations and Detection from distributed systems security literature, helps system defenders 
discover better RL-based defenses against attacks.

\begin{figure*}[t]
\centerline
{\includegraphics[width=0.995\linewidth]{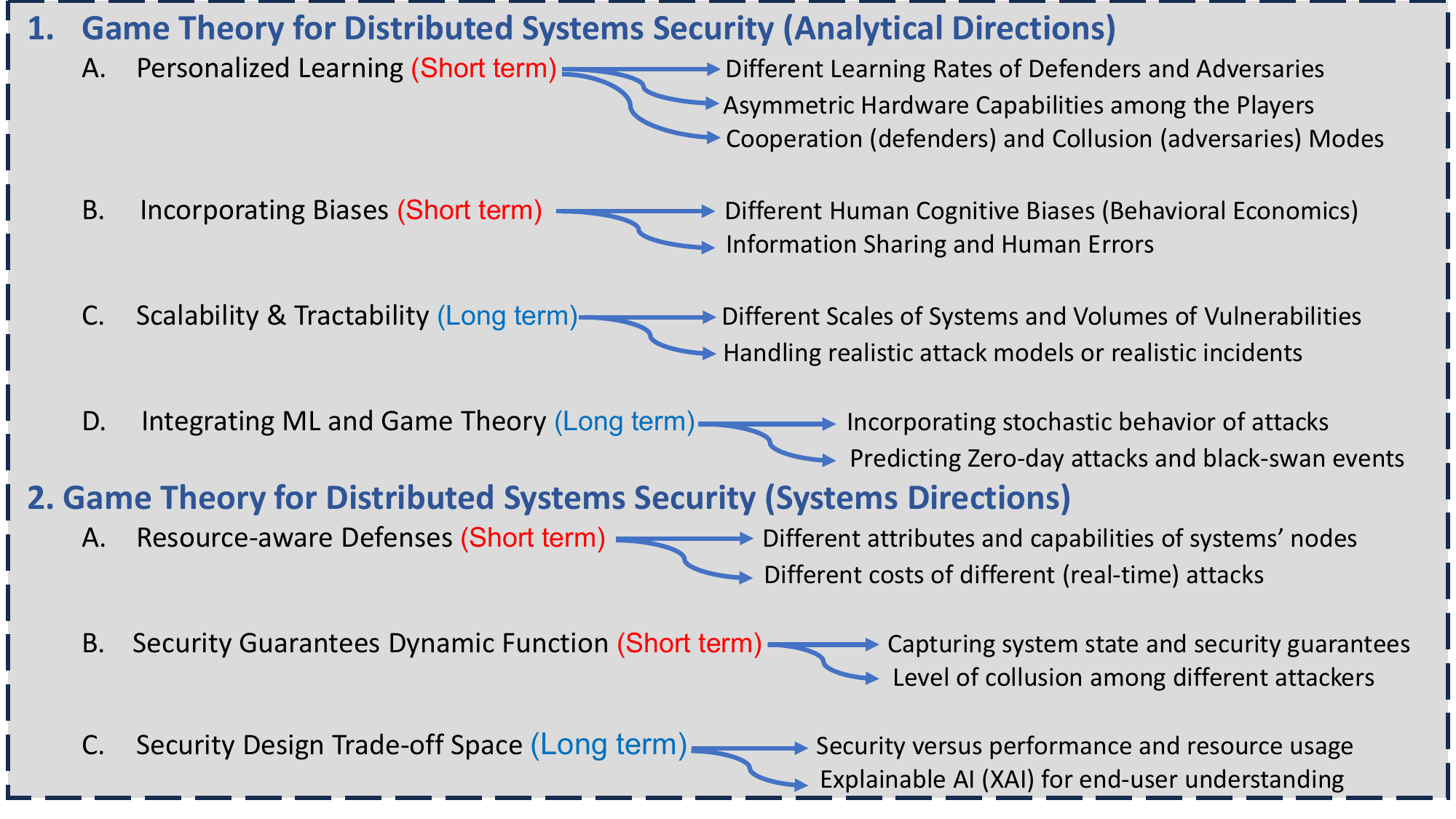}}
\vspace{-6mm}
\end{figure*}

    \noindent {\bf Integrated evaluation environments:} \textcolor{blue}{(Long time horizon)}
    The availability of an integrated evaluation environment (synonymous with testbed) would be important for the community to evaluate if we are making progress toward the hard security goals. These testbeds across application domains will have two parts, one would provide application-generic functionality and the other will be specific to the application domain. The desiderata for testbeds will be that they will allow for the injection of various types of attacks, the creation of different kinds of players (with the heterogeneous characteristics mentioned earlier in \textit{Analytical Directions}), and the measurement of a diversity of metrics of interest. 
    Such testbeds would be important for educating policymakers, in addition to their value in demonstrating research artifacts. The acquired education may help policymakers inform what kinds and degrees of security investments should be made to different parts of an interconnected system to reach a desired level of security. This will make our critical infrastructure more secure. 

In summary, our suggested research directions for each side (game-theory side and distributed systems security side), along  with integration of both sides can help in better development of tractable and practical methods for securing distributed systems, which would represent a consequential advancement for our technical community and our society broadly.

\section{CONCLUSION} 

Many of the critical infrastructure systems that we rely on as well as personal computing systems are structured as distributed systems. Take, for example, from the large end of the spectrum --- transportation, power grid, and financial infrastructures --- to the personal end of the spectrum --- cooperating personal computing and IoT devices.
As the attack surfaces for such systems become larger and the sophistication and the incentives for attacks against them increase, it is time to usher rigorous reasoning to secure such systems.  At the same time, the rigorous analytical foundations need to be made scalable and tractable to apply to these real-world large-scale applications under realistic resource  and timing constraints.

In this context, the technical themes of security of distributed systems and game theory applied to security have much to contribute to each other. We trust that the two vibrant communities will continue the process of working together for better securing distributed systems. In this article we have laid out a set of foundations that will serve as useful starting points for our journey plus a set of open research challenges that the community would hopefully take up. 

The challenges and research directions outlined in this vision paper
would help toward the ultimate goal of practical and secure distributed systems. 
Our broad vision is two-fold and applies to both distributed systems at work and in our personal lives. {\em First}, we will have a clear understanding of their security properties so that we can decide the level of trust that is warranted in each. {\em Second}, such reasoning can be done in a systematic manner, without having to perform one-off reasoning for each system.

\section{ACKNOWLEDGMENT}

The idea for this article was conceived of during the NSF SaTC PI Meeting in Arlington, Virginia in June 2022. We thank the attendees of the ``Game Theory and Distributed Systems Security'' breakout for their valuable feedback. We thank the NSF SaTC meeting organizers, William Enck, Heather Lipford, and Michael Reiter, and the NSF SaTC Program Managers, Jeremy Epstein, Nina Amla, and Daniela Oliveira, for enabling this session. This work was supported by the Army Research Lab (ARL) under Contract No. W911NF-2020-221 and National Science Foundation under grants CNS-202667, CNS-2134076, and CNS-2038986. Any opinions,
findings, and conclusions  expressed in
this material are those of the authors and do not necessarily
reflect the views of the sponsors.

\bibliographystyle{plain}
\bibliography{mustafa,bopardikar,murat}

\begin{IEEEbiography}{Mustafa Abdallah}{\,} is an Assistant Professor at the Purdue School of Engineering and Technology at  Indiana University-Purdue University Indianapolis (IUPUI). Dr. Abdallah got his Ph.D. from the Elmore Family School of 
Electrical and Computer Engineering at Purdue University in August 2022. 
Dr. Abdallah’s research interests include game-theory, human decision-making, and machine-learning with applications including cybersecurity, edge 
computing, and data science. 
Dr. Abdallah's research contribution is recognized by receiving the Purdue 
Bilsland Dissertation Fellowship and having many publications in top IEEE/ACM journals and conferences. He 
also was the recipient of
a M.Sc. Fellowship from Cairo University in 2013. 
Contact him at mabdall@iu.edu
\end{IEEEbiography}

\begin{IEEEbiography}{Saurabh Bagchi}{\,} is a Professor of Electrical and Computer Engineering and Computer Science at Purdue University. His research interest is in dependable computing and distributed systems. He is the founding Director of a university-wide resilience center at Purdue called CRISP (2017-present) and PI of the Army's Artificial Intelligence Innovation Institute (A2I2) (2020-25). Saurabh serves as the founder and CTO of a cloud computing startup, KeyByte. 
\end{IEEEbiography}

\begin{IEEEbiography}{Shaunak D. Bopardikar}{\,} is an Assistant Professor with the Electrical and Computer Engineering (ECE) Department at the Michigan State University (MSU). He received a Ph.D. degree in Mechanical Engineering from the University of California at Santa Barbara, USA, in 2010. From 2011-2018, he was a Senior/Staff Research Scientist with United Technologies Research Center at Berkeley, CA and East Hartford, CT, USA. He is a senior member of the IEEE, has over 75 refereed journal and conference publications, and holds 2 U.S. patented inventions. His recent recognitions include a National Science Foundation Career Award and an MSU College of Engineering's Withrow Teaching Excellence Award (ECE).  Contact him at \texttt{shaunak@egr.msu.edu}.
\end{IEEEbiography}

\begin{IEEEbiography}
{Kevin Chan} an Electronics Engineer at the U.S. Army Combat Capabilities Development Command, Army Research Laboratory, Adelphi, MD, USA. He is actively involved in research on network science, distributed analytics, and cybersecurity. He is the recipient of the 2021 IEEE Communications Society Leonard G. Abraham Prize and multiple best paper awards. He is a Senior Member of the IEEE and has  served as Co-Editor of the IEEE Communications Magazine—Military Communications and Networks Series.
\end{IEEEbiography}

\begin{IEEEbiography}
{Xing Gao} received the PhD degree in computer science from the College of William and Mary, Williamsburg, VA, USA, in 2018. He is currently an assistant professor with the Department of Computer and Information Sciences with the University of Delaware, Newark, DE, USA. From 2018-2020, he was an assistant professor with the Department of Computer Science, University of Memphis, Memphis, TN. His research interests include security, cloud computing, and mobile computing.
\end{IEEEbiography}

\begin{IEEEbiography}
    {Murat Kantarcioglu} is an Ashbel Smith Professor in the Computer Science Department and Director of the Data Security and Privacy Lab at The University of Texas at Dallas (UTD). He received a PhD in Computer Science from Purdue University in 2005 where he received the Purdue CERIAS Diamond Award for Academic excellence. He is also a faculty associate at Harvard Data Privacy Lab and a visiting scholar at UC Berkeley RISE Labs. Dr. Kantarcioglu's research focuses on the integration of cyber security, data science and blockchains for creating technologies that can efficiently and securely share and analyze data. He is also a fellow of AAAS (American Association for the Advancement of Science), IEEE, and a distinguished member of ACM.
\end{IEEEbiography}

\begin{IEEEbiography}{Congmiao Li, V,}{\,}
 received the B.S. degree in Computing (Computer Science) with minor in Mathematics and the M.S. degree in Electrical Engineering from National University of Singapore in 2009 and 2013, and the Ph.D. degree in Computer Engineering from University of California, Irvine (UCI) in 2020. She is a PostDoc at UCI. Her research interests include security and computer architecture.
\end{IEEEbiography}

\begin{IEEEbiography}
{Peng Liu, IV,}{\,} is the Raymond G. Tronzo, MD Professor of Cybersecurity, founding Director of the Center for Cyber-Security, Information Privacy, and Trust, and founding Director of the Cyber Security Lab at Penn State University. His research interests are in all areas of computer security. He has published over 350 technical papers, including numerous papers on top conferences and journals. His research has been sponsored by NSF, ARO, AFOSR, DARPA, DHS, DOE, AFRL, NSA, TTC, CISCO, and HP.  
He is the co-editor-in-chief of Journal of Computer Security and was an associate editor for IEEE TDSC. He is a recipient of the DOE Early Career Principal Investigator Award.  
He received his BS and MS degrees from the University of Science and Technology of China, and his PhD from George Mason University in 1999. He has advised or co-advised over 40 PhD dissertations to completion. 
\end{IEEEbiography}

\begin{IEEEbiography}
{Quanyan Zhu} is an associate professor at the Department of Electrical and Computer Engineering, New York University (NYU). He is an affiliated faculty member of the Center for Urban Science and Progress (CUSP) and Center for Cyber Security (CCS) at NYU. His current research interests include cognitive security, AI and game theory, cyber and physical system resilience, automation and robotics. He currently serves as the technical committee chair on security and privacy for IEEE Control Systems Society.  
\end{IEEEbiography}

\end{document}